\documentclass[aps,prl,superscriptaddress,showpacs,floatfix,twocolumn]{revtex4-2}

\usepackage[dvipdfmx]{graphicx}
\usepackage{mathrsfs}
\usepackage{float}
\usepackage{mediabb}
\usepackage{amsmath}
\usepackage{bm}
\usepackage[small,bf]{caption}
\usepackage[subrefformat=parens]{subcaption}
\usepackage{comment}

\newcommand{\beq}{\begin{equation}}
\newcommand{\eeq}{\end{equation}}
\newcommand{\beqa}{\begin{eqnarray}}
\newcommand{\eeqa}{\end{eqnarray}}
\newcommand{\bpr}{\begin{problem}}
\newcommand{\epr}{\end{problem}}
\newcommand{\bcent}{\begin{center}}
\newcommand{\ecent}{\end{center}}
\newcommand{\bfig}{\begin{figure}}
\newcommand{\efig}{\end{figure}}
\newcommand{\bpc}{\begin{picture}}
\newcommand{\epc}{\end{picture}}
\newcommand{\barr}{\begin{array}}
\newcommand{\earr}{\end{array}}
\newcommand{\bitm}{\begin{itemize}}
\newcommand{\eitm}{\end{itemize}}
\newcommand{\bright}{\begin{flushright}}
\newcommand{\eright}{\end{flushright}}
\newcommand{\bminip}{\begin{minipage}}
\newcommand{\eminip}{\end{minipage}}
\newcommand{\btab}{\begin{tabular}}
\newcommand{\etab}{\end{tabular}}

\newcommand{\nnb}{\nonumber}

\newcommand{\hiroshima}{Graduate School of Advanced Science and Engineering, Hiroshima University, Kagamiyama, Higashi-Hiroshima 739-8526, Japan}
\newcommand{\QUP}{International Center for Quantum-field Measurement Systems
for Studies of the Universe and Particles (QUP), KEK, Tsukuba, Ibaraki 305-0801, Japan}

\newcommand{\om}{\omega}

\newcommand{\al}{\alpha}
\newcommand{\be}{\beta}

\newcommand{\ve}{\varepsilon}
\newcommand{\la}{\lambda}

\newcommand{\B}{\langle}
\newcommand{\K}{\rangle}

\newcommand{\cm}[1]{\mbox{\hspace{#1 cm}}}
\newcommand{\mm}[1]{\mbox{\hspace{#1 mm}}}

\newcommand{\tSRPC}{$\mathrm{^tSRPC}$}

\begin{document}
\title{
Remote sensing of backward reflection from stimulated axion decay
}

\author{Kensuke Homma\footnote{corresponding author}}\affiliation{\hiroshima}\affiliation{\QUP}

\date{\today}

\begin{abstract}
We propose a method for remotely detecting backward reflection via induced decay of 
cold dark matter such as axion in the background of a propagating coherent photon field. 
This method can be particularly useful for probing concentrated dark matter streams 
by Earth's gravitational lensing effect. 
Formulae for the stimulated reflection process and the expected sensitivities in 
local and remote experimental approaches are provided for testing eV scale axion models 
using broad band lasers. 
The generic axion-photon coupling is expected to be explorable up to 
${\cal O}(10^{-12})$~GeV${}^{-1}$ and ${\cal O}(10^{-22})$~GeV${}^{-1}$ 
for the idealized local and remote setups, respectively.
%
\end{abstract}

\maketitle

\section{Introduction}
Despite that the topological nature in quantum Chromodynamics (QCD) rather naturally 
allows $CP$ violation, why is the $CP$-symmetry conserved in QCD ?
To resolve this strong $CP$ problem, Peccei and Quinn
have proposed the global $U(1)_{PQ}$ symmetry~\cite{Peccei:1977hh,Peccei:1977ur}.
As a result of the spontaneous PQ-symmetry breaking~\cite{Weinberg:1977ma, Wilczek:1977pj},
axion, a hypothetical particle, could appear as a kind of Nambu-Goldstone bosons 
with non-zero mass. If the energy scale for the symmetry breaking is much higher than that of
the electroweak symmetry, the coupling of axion to ordinary matter can be weak.
This invisible axion can be a candidate for cold dark matter~
\cite{Preskill:1982cy,Dine:1982ah,Abbott:1982af}.
Typically viable QCD axion models predict its mass much below the eV scale.
In contrast, there are several recent models which predict the mass range around eV
in the context of astrophobic QCD axion~\cite{DiLuzio:2017ogq,Badziak:2023fsc}
which are motivated to solve the anomalous cooling in several stellar environments
~\cite{Raffelt:2011ft,Giannotti:2015kwo,Giannotti:2017hny}.
Moreover, an axion-like particle (ALP) model relevant to inflation
predicts the eV range ALP~\cite{MIRACLE}.
Therefore, it is worth testing the eV scale QCD axion and ALP scenarios 
as in general as possible.

So far any of axion halo scopes have utilized static magnetic fields to induce axion decay
via the inverse Primakoff process~\cite{HaloScopes}. A giant magnet is typically not movable and thus
it is practically difficult to bring the searching system to a far region in the universe
where the axion density might be enhanced compared to the average density.
In concrete, a possibility that the dark matter density is locally concentrated
by the gravitational lensing from Earth has been evaluated~\cite{DMHair}.
The study predicts that
fine-grained streams of dark matter form the hair-like structure above the surface
of Earth and the typical distance to the roots of hairs is estimated as $\sim 10^9$ m
from Earth, where the concentration factor to the average density 
can reach $10^7-10^9$ depending on impact parameters of dark matter streams 
with respect to Earth~\cite{DMHair}.
If these hair-roots could exist around Earth
and be possibly stationary with respect to Earth's motion,
we are led to consider methods for detecting such remote stationary objects. 

\begin{figure}[H]
\begin{center}
\includegraphics[keepaspectratio,scale=0.42]{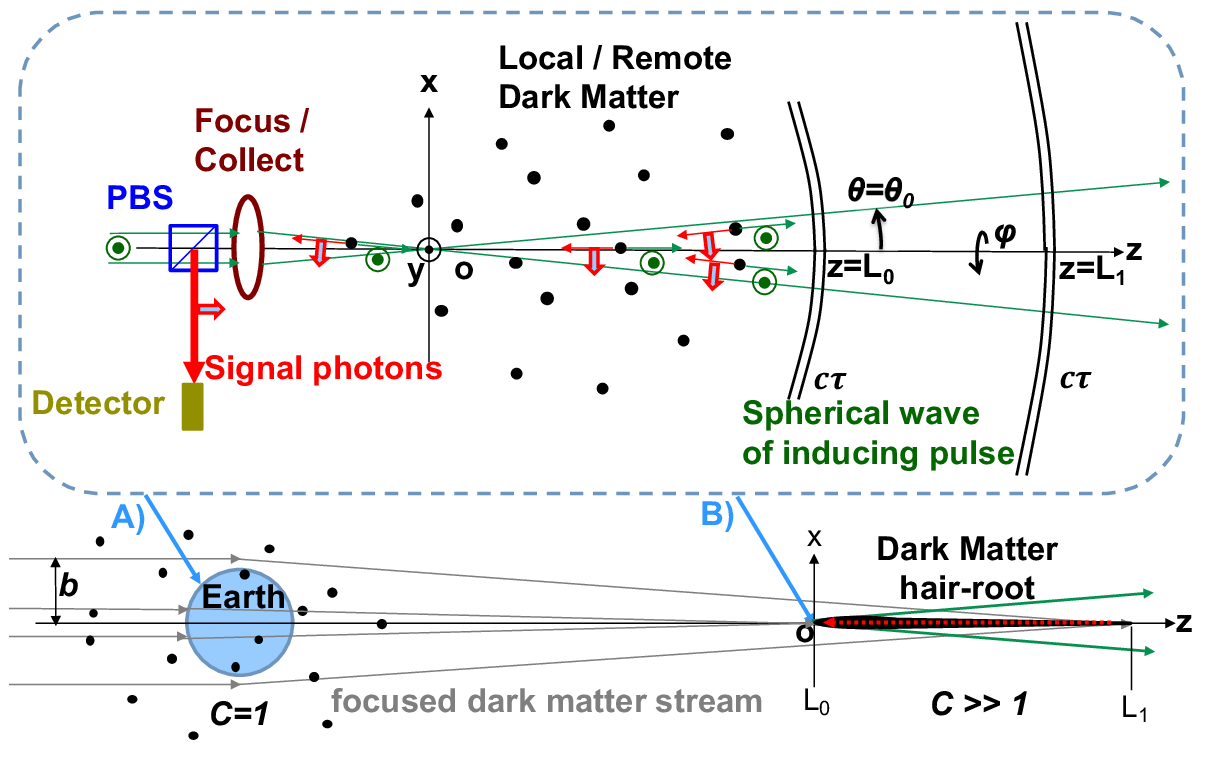}
\caption{
Concept for detecting backward reflection from stimulated decay of pseudoscalar dark matter 
such as axion. A coherent electromagnetic field to induce axion decay propagating through space
generates a spherical wavefront. The directions of signal photons from stimulated decay of
stationary particles must be opposite to those of the wave vectors (green arrows) 
along the spherical wavefront.
In the case of pseudosalar particles, the linear polarization directions between the inducing 
field (green $\odot$ aligned with the $y$-axis) and the decayed photons (red hollow arrows)
tend to be orthogonal to each other.
Since the wave vectors of the inducing field are perpendicular to the spherical wavefront, 
the decayed photons from cold dark matter reflect back to the light source. 
Consequently, the inducing spherical wave effectively acts as if a flying mirror,
providing a {\it light net } for capturing decayed photons at the source point.
As the net propagates at the velocity of light, a vast volume of cold dark matter
can be investigated almost instantaneously.
This concept is applicable to A) laboratory based experiments and B) remote sensing of
dark matter hair-roots due to the gravitational lensing effect by Earth with respect
to dark matter streams~\cite{DMHair}. $C$ expresses the local concentration factor 
relative to the average cold dark matter density around the Sun~\cite{Rho}. 
The other detailed explanations are found in the main text.
}
\label{Fig1}
\end{center}
\end{figure}
 
As illustrated in Fig.\ref{Fig1} 
we consider the stimulated decay process of axion in the coherent state
into two photons under an inducing photon beam in the coherent state. This process accompanies
generation of signal photons with different momenta and polarization states
from those of the inducing field in general. However, if axions are
non-relativistic and effectively at rest with respect to an observer,
due to energy-momentum conservation in the axion's rest frame,
the signal photons from stimulated decays must look 
as if reflection occurs from moving mirrors whose directions
are determined by the wave vectors of the inducing photons.
Thanks to the reversible nature of focusing and diverging wave vectors in 
the focused inducing beam,
the stimulated emission is expected to be fully collectable at the incident point of
the inducing beam.
In addition, since axion is a pseudoscalar particle,
the decayed two photons must have the characteristic correlation in the polarization states.
If either one photon is in a linearly polarization state,
the other one's polarization direction tends to be orthogonal.
Therefore, with respect to the polarization direction of the inducing field($y$-axis), 
selecting orthogonally polarized backward photons by a polarization beam splitter (PBS)
will help the discrimination from the trivial background process in the search.
In the following we formulate this stimulated reflection process and 
discuss the expected sensitivity to search for pseudoscalar-type dark matter
based on local and remote experimental approaches, A) and B), respectively.

\section{Invariant amplitude for stimulated decay of a coherent pseudoscalar field under a coherent photon field}
We discuss the generic interaction Lagrangian
between a pseudoscalar field $\sigma$ and two photons
\beq
-{\cal L} = gM^{-1} \frac{1}{4}F_{\mu\nu} \tilde{F}^{\mu\nu} \sigma
\eeq
where $g$ is a dimensionless coupling strength and $M$ is an energy scale
at which a symmetry breaking occurs with the photon field strength tensor
$F_{\mu\nu}$ and its dual tensor $\tilde{F}_{\mu\nu}$ whose definitions are given below.
The first order perturbation of S-matrix for the interaction Lagrangian
is expressed as
\beqa\label{eq_Smatrix}
S^{(1)} = \left(-\frac{1}{4}\frac{g}{M}\right) \frac{i^1}{1!} \int d^4x
N[F_{\mu\nu}(x)\tilde{F}^{\mu\nu}(x)\sigma(x)],
\eeqa
where $N$ denotes the normal-ordering of operators
with the following Fourier expansions of the relevant fields
\beq\label{eq_axion}
\sigma(x) \equiv
\int \frac{d^3 \bm{q}}{(2\pi)^3 2q^0}
(e^{-iqx}b_{\bm{q}} + e^{iqx}b^{\dagger}_{\bm{q}}),
\eeq
\beq\label{eq_Fmn}
F^{\mu\nu} \equiv
(-i)\int \frac{d^3 \bm{p}}{(2\pi)^3 2p^0}\Sigma_{\lambda=1,2}
(P^{\mu\nu}e^{-ipx}a_{\bm{p},\lambda} + \hat{P}^{\mu\nu}e^{ipx}a^{\dagger}_{\bm{p},\lambda}),
\eeq
and
\begin{align}\label{eq_FmnT}
\tilde{F}^{\mu\nu}
\equiv
\frac{1}{2}\ve^{\mu\nu\al\be}F_{\al\be} \mbox{\hspace{8.0cm}} \\ \nnb
=
(-i)\int \frac{d^3 \bm{p}}{(2\pi)^3 2p^0}\Sigma_{\lambda=1,2}
(\tilde{P}^{\mu\nu}e^{-ipx}a_{\bm{p},\lambda_p} + \hat{\tilde{P}}^{\mu\nu}e^{ipx}a^{\dagger}_{\bm{p},\lambda_p}), \mbox{\hspace{2.0cm}}
\end{align}
where we introduce the following momentum-polarization tensors as capitalized symbols
for an arbitrary four-momentum $p$ of the electromagnetic field with the polarization state $\la_p$:
\beqa\label{eq_Tensor}
P^{\mu\nu} &\equiv&        p^{\mu}e^{\nu}(\la_p)  -  e^{\mu}(\la_p)p^{\nu},
\\ \nnb
\hat{P}^{\mu\nu} &\equiv&  p^{\nu}e^{*\mu}(\la_p) -  e^{*\nu}(\la_p)p^\mu,
\eeqa
and
\beqa\label{eq_TensorT}
\tilde{P}^{\mu\nu} &\equiv& \frac{1}{2}\ve^{\mu\nu\al\be} \left( p_{\al}e_{\be}(\la_p) - e_{\al}(\la_p)p_{\be} \right),
\\ \nnb
\hat{\tilde{P}}^{\mu\nu} &\equiv& \frac{1}{2}\ve^{\mu\nu\al\be} \left( p_{\al}e^{*}_{\be}(\la_p) - e^{*}_{\al}(\la_p)p_{\be} \right)
\eeqa
with the Levi-Civita symbol $\ve^{ijkl}$.
%
The commutation relations between the photon creation and annihilation operators, 
$a_{\bm{p},\lambda}$ and $a^{\dagger}_{\bm{p},\lambda}$ are as follows
\beqa\label{eq_commutation}
[a_{\bm{k},\lambda},a^{\dagger}_{\bm{k}^{'},\lambda^{'}}] &=&
(2\pi)^3 2k^0 \delta^3(\bm{k}-\bm{k}^{'})\delta(\lambda-\lambda^{'}) \\ \nnb
&\equiv& \tilde{\delta}^3(\bm{k}-\bm{k}^{'})\delta(\lambda-\lambda^{'}), \\ \nnb
[a_{\bm{k},\lambda},a_{\bm{k}^{'},\lambda^{'}}] &=& [a^{\dagger}_{\bm{k},\lambda},a^{\dagger}_{\bm{k}^{'},\lambda^{'}}] = 0.
\eeqa
Henceforth we omit the polarization index $\lambda_p$ and
the sum over it because we require fixed beam polarizations in assumed experiments.
Substituting Eqs.(\ref{eq_axion}--\ref{eq_TensorT}) into Eq.(\ref{eq_Smatrix}),
we get
\begin{align}\label{eq_S1}
S^{(1)} = \left(-\frac{1}{4}\frac{g}{M}\right) \frac{i^1}{1} \int d^4x (-i)^2 
\times \mbox{\hspace{6cm}}\\ \nnb
      \int \frac{d^3 \bm{s}}{(2\pi)^3 2s^0}
      \int \frac{d^3 \bm{t}}{(2\pi)^3 2t^0}
      \int \frac{d^3 \bm{u}}{(2\pi)^3 2u^0} 
\times \mbox{\hspace{4.5cm}} \\ \nnb
N[
\bigl(
S_{\mu\nu}\tilde{T}^{\mu\nu}       e^{-i(s+t)x}a_{\bm{s}}a_{\bm{t}} +
S_{\mu\nu}\hat{\tilde{T}}^{\mu\nu} e^{-i(s-t)x}a_{\bm{s}}a^{\dagger}_{\bm{t}} +
\mbox{\hspace{3.6cm}} \\ \nnb
\hat{S}_{\rho\sigma}\tilde{T}^{\rho\sigma} e^{-i(t-s)x}a^{\dagger}_{\bm{s}} a_{\bm{t}} +
\mbox{\hspace{2mm}}
\hat{S}_{\rho\sigma}\hat{\tilde{T}}^{\rho\sigma} e^{i(s+t)x}a^{\dagger}_{\bm{s}}a^{\dagger}_{\bm{t}}
\bigr)
\mbox{\hspace{3.8cm}} \\ \nnb
\times \bigl(
e^{-iux}b_{\bm{u}} + e^{iux}b^{\dagger}_{\bm{u}}
\bigr)\mbox{\hspace{2mm}}
] \mbox{\hspace{7cm}}
\end{align}
where four momenta $s$, $t$ and $u$ correspond to those in
$F^{\mu\nu}$, $\tilde{F}^{\mu\nu}$ and $\sigma$, respectively.
Through Eqs.(\ref{eq_Tensor}-\ref{eq_TensorT}), the tensors defined by 
lowercase four-momenta are represented using their uppercase counterparts. 
Specifically, $S$ is used for $s$ and $T$ for $t$, 
similar to how $P$ is used for $p$ in Eqs.(\ref{eq_Tensor}-\ref{eq_TensorT}).
 
The definitions of the coherent state~\cite{Glauber} for a given
mean number of photons $N_{\bm{p}}$ with a three dimensional momentum $\bm{p}$
are summarized as follows:
\beq
|N_{\bm{p}}\K\K \equiv \exp\left(-N_{\bm{p}}/2 \right) \sum_{n=0}^{\infty}
\frac{N^{n/2}_{\bm{p}}}{\sqrt{n!}} |n_{\bm{p}}\K,
\label{coh_2}
\eeq
where $|n_{\bm{p}}\K$ is the normalized state of $n$ photons
\beq
|n_{\bm{p}}\K =\frac{1}{\sqrt{n!}}\left( a^\dagger_{\bm{p}}\right)^n |0\K,
\label{coh_3a}
\eeq
with the creation operator $a^\dagger_{\bm{p}}$
of photons that share a common momentum $\bm{p}$ and a common polarization state
over different number states.
The following relations on the coherent state
\beq
\B\B N_{\bm{p}} | N_{\bm{p}} \K\K = 1
\eeq
and
\beq
\B\B N_{\bm{p}}|n|N_{\bm{p}}\K\K =
\B\B N_{\bm{p}}|\left(a^\dagger_{\bm{p}} a_{\bm{p}}\right)|N_{\bm{p}}\K\K=N_{\bm{p}},
\label{coh_3e}
\eeq
give us basic properties with respect to the creation and annihilation operators
to the coherent state:
\beq\label{eq_enhancement}
a_{\bm{p}}|N_{\bm{p}}\K\K =\sqrt{N_{\bm{p}}}|N_{\bm{p}}\K\K ,\quad\mbox{and}\quad
\B\B N_{\bm{p}}|a^\dagger_{\bm{p}} =\sqrt{N_{\bm{p}}}\B\B N_{\bm{p}}|.
\eeq
For the coherent state of a pseudoscalar field we apply the same properties
to the annihilation and creation operators $b_{\bm{p}}$ and $b^{\dagger}_{\bm{p}}$, 
respectively,
\beq\label{}
b_{\bm{p}}|{\cal N}_{\bm{p}}\K\K =\sqrt{{\cal N}_{\bm{p}}}|{\cal N}_{\bm{p}}\K\K ,\quad\mbox{and}\quad
\B\B {\cal N}_{\bm{p}}|b^\dagger_{\bm{p}} =\sqrt{{\cal N}_{\bm{p}}}\B\B {\cal N}_{\bm{p}}|
\eeq
with 
\beq
\B\B {\cal N}_{\bm{p}} | {\cal N}_{\bm{p}} \K\K = 1.
\eeq 

We consider the stimulated decay process of a pseudoscalar field
in the coherent state $|{\cal N}_{\bm{q}}\K\K$ into two photons
via the four-momentum exchange: $q \rightarrow p_i + p_s$ 
under an inducing fields $|N_{\bm{p}_i}\K\K$ in the coherent state of photons
accompanying generation of a signal photon $p_s$.
The initial and final states, respectively, are thus defined as follows:
\beqa
|\Omega\K &\equiv& |{\cal N}_{\bm{q}}\K\K |N_{\bm{p}_i}\K\K |0\K \quad \mbox{and}\\ \nnb
\B \Omega^{'}| &\equiv&  \B\B {\cal N}_{\bm{q}}|\B\B N_{\bm{p_i}}| \B 1_{\bm{p_s}}| 
=\B\B \Omega | a_{\bm{p_s}}.
\eeqa

The transition amplitude is then expressed as follows
\begin{align}
\label{eq_S1org}
\B \Omega^{'}| S^{(1)} |\Omega\K
= \left(-\frac{1}{4}\frac{g}{M}\right) \frac{i^1}{1} \int d^4x (-i)^2 
\times \cm{7} \\ \nnb
\int \frac{d^3 \bm{s}}{(2\pi)^3 2s^0} \int \frac{d^3 \bm{t}}{(2\pi)^3 2t^0} 
\B\B N_{\bm{p_i}}| \B 1_{\bm{p_s}}|
\mbox{\hspace{7cm}} \\ \nnb
\bigl(
S_{\mu\nu}\tilde{T}^{\mu\nu}       e^{-i(s+t)x}a_{\bm{s}}a_{\bm{t}} +
S_{\mu\nu}\hat{\tilde{T}}^{\mu\nu} e^{-i(s-t)x}a^{\dagger}_{\bm{t}}a_{\bm{s}} +
\mbox{\hspace{5.5cm}} \\ \nnb
\hat{S}_{\rho\sigma}\tilde{T}^{\rho\sigma} e^{-i(t-s)x}a^{\dagger}_{\bm{s}} a_{\bm{t}} +
\hat{S}_{\rho\sigma}\hat{\tilde{T}}^{\rho\sigma} e^{i(s+t)x}a^{\dagger}_{\bm{s}}a^{\dagger}_{\bm{t}}
\bigr)
|N_{\bm{p_i}}\K\K |0\K \times \mbox{\hspace{4.3cm}} \\ \nnb
\int \frac{d^3 \bm{u}}{(2\pi)^3 2u^0}
\B\B {\cal N}_{\bm{q}}|
\bigl(
e^{-iux}b_{\bm{u}} + e^{iux}b^{\dagger}_{\bm{u}}
\bigr)
|{\cal N}_{\bm{q}}\K\K .
\mbox{\hspace{6cm}}
\end{align}
Results of photon annihilation-creation operators to the given states are 
summarized as follows
\beqa\label{eq_a}
a_{\bm{k}}|N_{\bm{p_i}}\K\K |0\K 
= \tilde{\delta}^{3}(\bm{k}-\bm{p_i})\sqrt{N_{\bm{p_i}}} |N_{\bm{p_i}}\K\K |0\K
+ |N_{\bm{p_i}}\K\K a_{\bm{k}}|0\K \mm{6} \\ \nnb
= \tilde{\delta}^{3}(\bm{k}-\bm{p_i})\sqrt{N_{\bm{p_i}}} |N_{\bm{p_i}}\K\K |0\K.
\mm{26}
\eeqa
\beq\label{eq_aa}
a_{\bm{l}}a_{\bm{k}}|N_{\bm{p_i}}\K\K |0\K = \tilde{\delta}^{3}(\bm{k}-\bm{p_i})\sqrt{N_{\bm{p_i}}} |N_{\bm{p_i}}\K\K a_{\bm{l}}|0\K = 0 \cm{4}
\eeq
because of $\bm{l}\ne \bm{k}$.
\beqa\label{eq_ad}
\B\B N_{\bm{p_i}}| \B 1_{\bm{p_s}}|a^{\dagger}_{\bm{k}} =
\B\B N_{\bm{p_i}}| \B 1_{\bm{p_s}}| \tilde{\delta}^{3}(\bm{k}-\bm{p_i})\sqrt{N_{\bm{p_i}}} +
\cm{1} \\ \nnb
\B\B N_{\bm{p_i}}| \B 0| \tilde{\delta}^{3}(\bm{k}-\bm{p_s}).
\mm{27}
\eeqa
\beqa\label{eq_adad}
\B\B N_{\bm{p_i}}| \B 1_{\bm{p_s}}|a^{\dagger}_{\bm{k}} a^{\dagger}_{\bm{l}} &=& 
\B\B N_{\bm{p_i}}| \B 0| \tilde{\delta}^{3}(\bm{k}-\bm{p_i})\sqrt{N_{\bm{p_i}}}
\tilde{\delta}^{3}(\bm{l}-\bm{p_s})
\mm{6}
\\ \nnb
&+& \B\B N_{\bm{p_i}}| \B 0| \tilde{\delta}^{3}(\bm{k}-\bm{p_s})
\tilde{\delta}^{3}(\bm{l}-\bm{p_i})\sqrt{N_{\bm{p_i}}}
\eeqa
because of $\bm{l}\ne \bm{k}$.
Combination of Eqs.(\ref{eq_a}) and (\ref{eq_ad}) gives
\beqa\label{eq_ada}
\B\B N_{\bm{p_i}}| \B 1_{\bm{p_s}}|a^{\dagger}_{\bm{l}} a_{\bm{k}}|N_{\bm{p_i}}\K\K |0\K 
= 
\tilde{\delta}^{3}(\bm{l}-\bm{p_s})
\tilde{\delta}^{3}(\bm{k}-\bm{p_i})\sqrt{N_{\bm{p_i}}}.
\mm{6}
\eeqa
By substituting Eqs.(\ref{eq_aa}),(\ref{eq_adad}) and (\ref{eq_ada})
into Eq.(\ref{eq_S1org}), we obtain the invariant amplitude kinematically consistent
with the stimulated decay process: $q \rightarrow p_i + p_s$ as follows
\beqa
\B \Omega^{'}| S^{(1_{dcy})} |\Omega\K
=(-i) (2\pi)^4 \delta^4(q-p_i-p_s) {\cal M}_i
\eeqa
with
\beq\label{eq_decayamp}
{\cal M}_i \equiv
2\sqrt{N_{\bm{p_i}}} \sqrt{{\cal N}_{\bm{q}}} \left(-\frac{1}{4}\frac{g}{M}\right)
\hat{P_i}_{\rho\sigma}\hat{\tilde{P_s}}^{\rho\sigma}
\eeq
and
\beqa\label{eq_vertex}
\hat{P_i}_{\rho\sigma}\hat{\tilde{P_s}}^{\rho\sigma} 
= \ve^{\mu\nu\alpha\beta} p_{i \mu} p_{s \alpha} e^*_{i \nu} e^*_{s \beta}
= -2 \om^2_i \cos\Phi
\eeqa
where $\om_i(=\om_s)=m/2$ with $\om_i \equiv p^0_i$ and $\om_s \equiv p^0_s$ is the photon energy 
of the inducing field 
corresponding to half of the mass of a pseudoscalar particle, $m$, and
$\Phi$ is a relative angle between the direction of the electric field component of
the linear polarization of a decayed photon with respect to
that of the magnetic field component of the linearly polarized inducing photons.

\section{Differential decay rate}
Given an invariant amplitude ${\cal M}$
for the spontaneous decay rate of a rest particle 
with its four momentum $p_0$ and mass $m$ into two photons : 
$p_0 \rightarrow p_1 + p_2$, the differential decay rate $d\Gamma$ is expressed as
\beq
d\Gamma = 
\frac{1}{2m}\frac{1}{2!}
\int\frac{d^3\bm{p_1}}{(2\pi)^3 2\om_1}
\int\frac{d^3\bm{p_2}}{(2\pi)^3 2\om_2}
(2\pi)^4 \delta^4(p_1+p_2-p_0)|{\mathcal{M}}|^2
\eeq
where we introduced the factor of $1/2!$ because the two photons with energy $\om_1$ and
$\om_2$ are identical particles.
Then the differential rate with respect to a solid angle of $p_1$ is eventually expressed as
\beq\label{eq_dGdO}
\frac{d\Gamma}{d\Omega_1} = \frac{|{\mathcal{M}}|^2}{128\pi^2 m}.
\eeq
\section{Stimulated backward reflection yield}
By substituting Eq.(\ref{eq_vertex}) into Eq.(\ref{eq_decayamp}), 
we can express the invariant amplitude for the stimulated decay process as
\beq\label{eq_decayampcos}
{\cal M}_i = \sqrt{N_{\bm{p_i}}} \sqrt{{\cal N}_{\bm{q}}} \left(\frac{g}{M}\right)
\om^2_i\cos\Phi.
\eeq
In order to equate ${\cal M}_i$ in Eq.(\ref{eq_decayampcos}) with ${\cal M}$ in Eq.(\ref{eq_dGdO}),
we must require that momentum and polarization of either one of decayed photons coincides with
those of an inducing beam. This requirement originates from the delta function
in the commutation relation in Eq.(\ref{eq_commutation}) as explicitly shown in the derivation
of the amplitude.
If a pseudoscalar particle exists at rest, $p1$ and $p_2$ photons must be emitted back-to-back
to each other with equal energies.
Therefore, if $p_1$ photon in the spontaneous decay process coincides with
$p_i$ within a range of the inducing beam, the stimulated emission of $p_2$, that is,
the signal photon $p_s$ must be exactly backward with respect to the $p_1$ direction.
For later convenience, we define the induced decay rate as
\beq
\frac{d\Gamma_i}{d\Omega} \equiv  N_i {\cal N}_q \frac{d\Gamma_0}{d\Omega}
\eeq 
with 
\beq\label{eq_dG0dO}
\frac{d\Gamma_0}{d\Omega} \equiv \frac{1}{128\pi^2 m} \left(\frac{g}{M}\right)^2 \om^4_i \cos^2\Phi
\eeq
in order to factor out the field strength parameters. 

Based on the above picture, we evaluate the stimulated reflection yield.
For a laser field a Gaussian beam as a function of spacetime point $(t, x, y, z)$ is 
typically used to express a focused field~\cite{Yariv}.
We adopt the spherical wave propagation which is interpreted as the approximated
wave propagation of a Gaussian beam at a far distance from the focal point~\cite{Yariv}.
Since in this case the wavefront simply form a sphere, the entire
solid angle of the inducing laser propagating from and to the focal point
is available as the inducing decay area. 
Thus backward signal photons are, in principle, 
fully collectable at the incident point of the inducing field
if dark matter particles are at rest with respect to the incident point,
wherever signal photons are generated. 
This is thanks to the reversible nature of the coherent light propagation.

We consider the case in the upper part of Fig.\ref{Fig1}
where a pulsed inducing field with duration of $\tau$ 
is focused once and the searching beam is generated from the focal point.
The central pulse position $z$ of the inducing field is moving with the velocity of light $c$
along the $z$-axis in free space which is occupied by coherent axions
with the number density $\rho_a$.
The stimulated reflection yield is then parametrized as follows
\beq
{\cal Y} = \int^{L_1/c}_{L_0/c} dt \int \rho_i dV_i \int \rho_a dV_i
           \int^{\Omega_i}_0 \frac{d\Gamma_0}{d\Omega} d\Omega,
\eeq
where $V_i$ specifies the volume scanned by the inducing field
while it propagates over the searching region from $z=L_0$ to $L_1$
with the solid angle
$\Omega_i = \int^{\theta_0}_0 d\theta \sin\theta \int^{2\pi}_0 d\phi = 2\pi (1-\cos\theta_0)$
defined by the divergence angle $\theta_0$ determined by geometrical optics.
The number density of the inducing field $\rho_i$ is parametrized as
$\rho_i(z) = \frac{N_i}{c\tau z^2 \Omega_i}$ as a function of $z$
reflecting the situation that the pulse center position moves along the z-axis 
with the surface area of $z^2 \Omega_i$ at $z$.
As for the volume integral on the axion density, we express it as
$
{\cal N}_q \equiv \int \rho_a dV_i = 
\int^{\infty}_{-\infty} dz^{'} \rho_a c\tau {z^{'}}^2 \Omega_i \delta(z^{'}-z), 
$
where the delta function requires local overlaps with inducible ranges 
along the propagation of the inducing field.
The stimulated reflection yield while the inducing field propagates is eventually estimated as
\begin{align}\label{eq_Y}
{\cal Y} 
= 
\int^{L_1/c}_{L_0/c} dt \int^{L_1}_{L_0} dz z^2 \int^{\theta_0}_{0} d\theta \sin\theta \int^{2\pi}_{0} d\phi
\frac{N_i}{c\tau z^2 \Omega_i} \times \cm{6} \\ \nnb
\int^{\infty}_{-\infty} dz^{'} \rho_a c\tau {z^{'}}^2 \Omega_i \delta(z^{'}-z)
\frac{d\Gamma_0}{d\Omega} \Omega_i \cm{6} \\ \nnb
=
\frac{L_1-L_0}{c} N_i \cdot \rho_a \frac{2}{3}\pi (L^3_1-L^3_0) \Omega_i  \cdot \frac{d\Gamma_0}{d\Omega} \Omega_i
\cm{7.5} \\ \nnb
=
\frac{\pi}{48} \left[\frac{1}{m} \left(\frac{g}{M}\right)^2 \om^3_i \right] \times \cm{10.2} \\
\left(\frac{L_1-L_0}{c}\frac{c}{\lambda_i}\right) \cos^2 \Phi N_i 
\rho_a (L^3_1-L^3_0) (1-\cos\theta_0)^2 \cm{6} \nnb
\end{align}
where the factor enclosed by $\left[\quad\right]$ partially keeps natural units
while $\om_i$ out of $\om^4_i$ in Eq.(\ref{eq_dG0dO}) is replaced with frequency 
using the wavelength of the inducing field $\lambda_i$.
It is important to note that the stimulated reflection yield eventually does not depend on 
whether the inducing field is provided as a pulsed wave or a continuous wave.

\section{Searching scenarios and the sensitivity projections}
\subsection{Searching scenarios}
We provide sensitivity evaluations in the following two experimental approaches:
A) laboratory based experiment targeting on the mean number density of axions around the Sun 
as cold dark matter~\cite{Rho},
B) remote detection aiming at the locally concentrated number density 
in a hair-root due to the gravitational lensing effect by Earth.
We thus introduce the local concentration factor $C$ as follows
\beq
\rho_a = C \frac{\rho_0}{m [\mbox{eV}]}
\eeq
with
$\rho_0 = 0.4 \times 10^9 \quad [\mbox{eV/cm}^{3}]$~\cite{Rho}.
In the standard halo model (SHM) for dark matter
the local velocity on Earth is traditionally assumed to follow 
the Maxwellian velocity distribution. 
On the other hand, it is discussed that the lensing effect is diminished
if the velocity dispersion is large in SHM~\cite{SofueLens}.
There might also be a fraction of cold dark matter whose
velocity dispersion is quite small~\cite{Sikivie95}.
In this case the lensing effect can be significant.
Therefore, we parametrize the generic concentration factor as follows:
\beq
C \equiv f_{\rho} \mathscr{M}, 
\eeq
where $f_{\rho}$ is the reduction factor due to the picture that
the energy density of generic streams $\rho_s$ is a fraction of averaged density $\rho_0$,
and $\mathscr{M}$ is density magnification purely due to Earth's lensing effect 
with dispersionless incident velocity. 
For instance, in SHM not assuming fine-grained streams, 
$\mathscr{M} \sim 10^9$ for a stream velocity of 17~km/s and 
$f_{\rho} \propto (\delta v / \sigma_v)^3$ 
with a velocity range $\delta v$ and velocity dispersion of SHM $\sigma_v$
gives $C = \mathcal{O}(10)$~\cite{SofueLens}.
On the other hand, if streams are composed of fine-grained streams,
the reduction factor would significantly depend on the numerical 
calculation on the fine-grained structure of cold dark matter haloes~\cite{FineGrain}.

In the following subsections, for A) and B), respectively, we provide the evaluations of
the density concentration factors $C$ and the signal acceptance factors
$A_{cc}$, assuming a signal collector such as a parabolic mirror or a lens
system with a radius $R_c$. We then present sensitivity projections
based on the common formulation.

\subsection{A) Remote sensing on the ground}
SHM is a natural assumption to evaluate the local energy density.  
In this case we can simply take $C=1$.

We assume that dark matter particles are moving with the escape velocity 
from the Galaxy $\sim 300$ km/s~\cite{Rho} at maximum
to any observation points, that is, the maximum relative velocity of 
$\beta = {\cal O}(10^{-3})$ in units of the velocity of light $c$.
In order to allow deviations of induced emission angles from the exact backward direction
due to the slight transverse Lorentz boost,
the radius of the signal collector, $R_c$, must satisfy 
$R_c \sim \beta L_1$ with the depth of the searching volume $L_1$.
Therefore, if $R_c=1$~m is assumed, the searchable depth is limited to $L_1 \sim 10^3$~m.
In this case the acceptance factor to all of signal photons can be $A_{cc} \sim 1.0$.

\subsection{B) Remote sensing in space}
We specifically target a possible dark matter (DM) source in the S1 stream from Cygnus, which
has been discovered in data from the Sloan Digital Sky Survey (SDSS) \cite{SDSS} and 
the Gaia satellite \cite{Gaia}. Since the progenitor of S1 is a dwarf galaxy 
with a sufficiently large total mass $M_0 \sim 10^{10} M_{\odot}$ and its infall time $T_{in} \ge 9\times 10^9$~yr, 
the stream may contain significant DM components.
The average Galactic position of the stars in the S1 stream source is 
$\mbox{(X, Y, Z)}= (8.9, 0.6, 2.5)$~kpc with the dispersion $(1.6,1.4,1.9)$~kpc, 
while the Solar position is $(8.2, 0, 0.014)$~kpc~\cite{S1_46,S1_47}, and
the distance between them is thus $D=2.7$~kpc.
So the stream of DM associated with S1 is expected to constantly hit the Solar System
and there is $\mathcal{O}(1)$ probability for the DM stream to appear 
in the Solar System~\cite{S1DM}.
Therefore, if a satellite trailing the Earth's orbit is designed to constantly face
the opposite direction of the S1 stream, 
the probability of finding a stationary hair-root caused by the S1 stream can be unity.
Thus, in principle, the satellite with a proper shield against sunlight is expected 
to be able to observe the hair-root in a stationary condition
for close to 6 months per year.

Since the density of the DM component cannot easily be measured from the stellar stream,
we model the DM stream density $\rho(v)$ at the incident plane in front of the Earth lens
for a given incident stream velocity $v$
based on the following simple picture with conservative assumptions.
Suppose a stream source with the total energy $E(t)$ which
isotropically emits dark matter streams as a function of time $t$ 
in the process of infall with $E(0) = M_0$.
For a short time period compared to the infall time scale $T_{in}$,
the emission rate can be approximated with a differential equation
$dE = -\tau^{-1} dt E$ as a result of emission
of dark matter above the escape velocity $v_{esc}$.
We assume $\tau \sim T_{in}$ because the age of the Universe may be similar to 
the infall time scale.
In this case the emission rate $r_e$ observed at the incident plane of the Earth lens
is approximated as
\beq
r_e \sim \tau^{-1} M_0 \frac{\pi R^2_E}{4\pi D^2}
\eeq
where the radius of Earth $R_E$, $M_0 = 10^{10} M_{\odot}$, and $\tau = 10^{10}$~yr are used.
Given the incident rate, the local energy density flowing into a cylindrical
volume with the base area corresponding to the Earth's cross section
in short time duration $\Delta t$ is expressed as
\beq
\rho(v) = \frac{r_e \Delta t}{v \Delta t \pi R^2_E} = \frac{\tau^{-1} M_0}{4\pi D^2 v}.
\eeq
The distance between the S1 source and the Sun, $D=8.182 \times 10^{19}$~m,
is much longer than $R_E=6.371 \times 10^6$~m, resulting in
an incident angle of $\mathcal{O}(10^{-13})$ rad.
Since the emission source is not necessarily limited to the central position of the S1 stream, 
incident angles toward the Earth lens can be three-dimensional given the kpc-scale source size. 
However, because we assume that the satellite constantly faces the central position, 
we may expect that almost parallel streams appear within the small solid angle, effectively reducing
the problem to one dimension.
Therefore, the assumption of a cylindrical volume for the single stream is indeed valid. 
We note, however, this viewpoint does not preclude the appearance of plural hair-roots around Earth,
which the satellite does not face.
As a conservative approach,
we assume the velocity distribution of DM follows SHM at the S1 source and
the satellite receives a DM stream within the narrow solid angle.
In this case the dark matter energy density at the incident region around Earth
is expressed with an effectively one-dimensional velocity dispersion as follows
\beq\label{eq_rhos}
\rho_s  = \int_0^{\infty} dv
\frac{\rho(v)}{\sqrt{2\pi\sigma^2_v}} \exp\left(-\frac{(v-v_{esc})^2}{2\sigma^2_v}\right)
\eeq
where $\rho(v)$ is integrated with the Gaussian weight assuming
the escape velocity $v_{esc}=520$~km/s and the standard deviation $\sigma_v=46$~km/s
as discussed in~\cite{S1DM}.
We thus can evaluate $f_{\rho} = \rho_s/\rho_0 = 0.002$ from Eq.(\ref{eq_rhos}), which is conservative
compared to $f_{\rho} = 0.1$ with $\rho_0=0.5$ GeV/cm${}^3$ in \cite{S1DM}.
As we discuss in the next paragraph, by requiring a long focal point,
we can clearly separate the S1-originating hair-root from those caused by
the local DM halo following SHM with $v_{esc} < 300$~km/s.

The length of the hair-root is estimated by using Eq.(36) in \cite{DMHair} where
the focal length $F$ for dark matter streams incident with velocity $v$
as a function of the impact parameter $b$ is expressed as
\beq
F(b) = \frac{\xi R^2_E}{3}\left(1+\frac{b^2}{4R^2_E}\right)
\eeq
with $\xi \equiv v^2/(G M_E) = 6.8 \times 10^{-4}$~m${}^{-1}$ for $v=v_{esc}=520$~km/s,
Earth's mass $M_E$, and the gravitational constant $G$. 
By changing $b$ from 0 to $R_E$,
the expected hair-root range from $F(0)= 9.20 \times 10^9$~m to 
$F(R_E) = 11.5 \times 10^9$~m is obtained. 
It is known that magnification for massive particles incident at $b \sim 0$ becomes
$\mathscr{M} \sim 10^9$ whereas particles incident at $b \sim R_E$ is $\mathscr{M} \sim 10^7$~\cite{DMHair}.
We conservatively take a constant magnification factor $\mathscr{M}\sim 10^7$ over the hair-root range.
So the concentration factor is eventually estimated as $C=f_{\rho} \mathscr{M} \sim 2 \times 10^4$.

Given $F(R_E)=5 \xi R^2_E/12$, the indent angle of dark matter streams
at the focal point is estimated as $\sim R_E/F(R_E)=12/(5 \xi R_E) = 5.5 \times 10^{-4}$ rad.
Therefore, the transverse velocity in units of $c$ in a focused hair-root 
with respect to the propagation direction of the inducing beam
is estimated as $\beta = 9.6 \times 10^{-7}$ given the incident velocity $v=520$~km/s.
It is practically impossible to collect all of signal photons in this case.
By assuming the satellite location at $F(0)$,
the acceptance factor of the signal collector with the radius $R_c=1$~m is estimated 
as $A_{cc} \sim [ R_c / \{ \beta ( F(R_E)-F(0) ) \} ]^2 = 2.0 \times 10^{-7}$.

\subsection{Sensitivity projections}
To provide the inducing field in the coherent state, 
we consider broadband pulsed lasers,
since the target axion in this paper is in the eV mass range and the axion mass is
not known.
In terms of the rate of the induced yield of backward reflection signals,
as we discussed,
there is no physical difference on the choice of a pulsed or continuous laser, that is,
the average power is the crucial parameter to discuss the yield.
However, the choice of pulsed lasers is beneficial when considering atomic background 
processes, because the de-excitation time scale in atoms is generally long, 
whereas the time scale of the induced decay of axions is prompt.
Therefore, the arrival timing of the reflected photons contains information 
that could potentially classify the de-excitation and the prompt decay components.
With pulsed lasers, we can utilize time-of-flight information synchronized with 
the pulse injection timing, and the pulsed laser system also enables simultaneous 
background photo-counting measurements during pulse intervals.

In order to scan a wide mass range simultaneously, for instance, 
pulsed lasers from an optical comb~\cite{OpticalCom} would be optimal.
In the following evaluation we assume the availability of an optical comb
covering $0.35$~$\mu$m - $4.4$~$\mu$m 
\cite{OpticalCom} by assuming a flat wavelength spectrum for simplicity.
Although the cited comb laser can generate up to 30 mW average power 
at a 48 MHz repetition rate in the visible range, 
in order to scale the sensitivity projections, for simplicity, 
we assume the availability of a 1~W-class comb laser in the near future.
Indeed, as another example of a relatively less broadband laser compared to comb lasers, 
a Ti:Sapphire pulse laser with
2~J / 20~fs duration pulse at a 10~Hz repetition rate, corresponding to 
20~W average power (100 TW peak power), is practically used for 
a real experiment~\cite{Challenge}.  
The average power of 1~W can also be interpreted
as 1 J / $10^7$ pulses operated at a 10 MHz repetition rate if a high repetition rate
laser such as a typical comb laser is considered.
Thus the parameter 1~J / pulse at a 1~Hz repetition rate 
can represent several types of realistic broadband pulsed lasers
achievable with the current laser technology.
Therefore, this laser parameter will be commonly assumed for the following evaluations.
We also note that in the case of insufficient average laser power, the pulse statistics, 
namely, the data acquisition time can compensate for the shortfall.
Thus the number of shots will be conservatively assumed to be $10^6$
for 1~J / pulse at 1~Hz.

In the both approaches A) and B), we consider measuring the arrival time distribution of 
signal photons with the orthogonally rotated linear polarization state to the
linear polarization direction of the inducing laser
by requiring the same photon energy as that of the inducing laser.
We assume a single photon sensitive device such as a photomultiplier
whose random thermal noise rate can be typically suppressed at $r_{thn} = 1$~Hz.
The overall signal collection factor including the detection efficiency is 
assumed to be $\epsilon = 0.1 A_{cc}$.
The repetition rate of the inducing pulse laser is assumed to be $f=1$~Hz. 
Within the scanning time window of $(L_1-L_0)/c$ per shot, 
the accidental coincidence rate between the thermal noise rate
and the laser repetition rate is expressed as $R_{bkg} \equiv 2 r_{thn} f (L_1-L_0)/c$. 
We thus can discuss a null hypothesis as Gaussian fluctuations after subtracting 
the total background statistics $T R_{bkg}$ with the data acquisition time $T=10^6$~sec
from the observed number of photons consistent with the signal feature.
In order to evaluate the sensitivity,
we require that the observed yield $f T \epsilon {\cal Y}$ based on Eq.(\ref{eq_Y}) 
is greater than five standard deviations of the subtraction-associated fluctuation
over the total statistics, that is, $\delta N_{bkg} = 5\sqrt{2 T R_{bkg}}$.
Therefore, the upper limit of the coupling strength when no signal photon is observed
is express as
\beq\label{eq_goM}
\frac{g}{M} = \sqrt{
\frac {\delta N_{bkg}}
 {f T \epsilon \frac{\pi \om^3_i}{48 m}\left(\frac{L_1-L_0}{\lambda_i}\right) N_i
 \rho_a (L^3_1-L^3_0) (1-\cos\theta_0)^2 
 }
}
\eeq
by requiring the orthogonal polarization state $\cos \Phi = 1$.

\begin{figure}
\begin{center}
\includegraphics[keepaspectratio,scale=0.45]{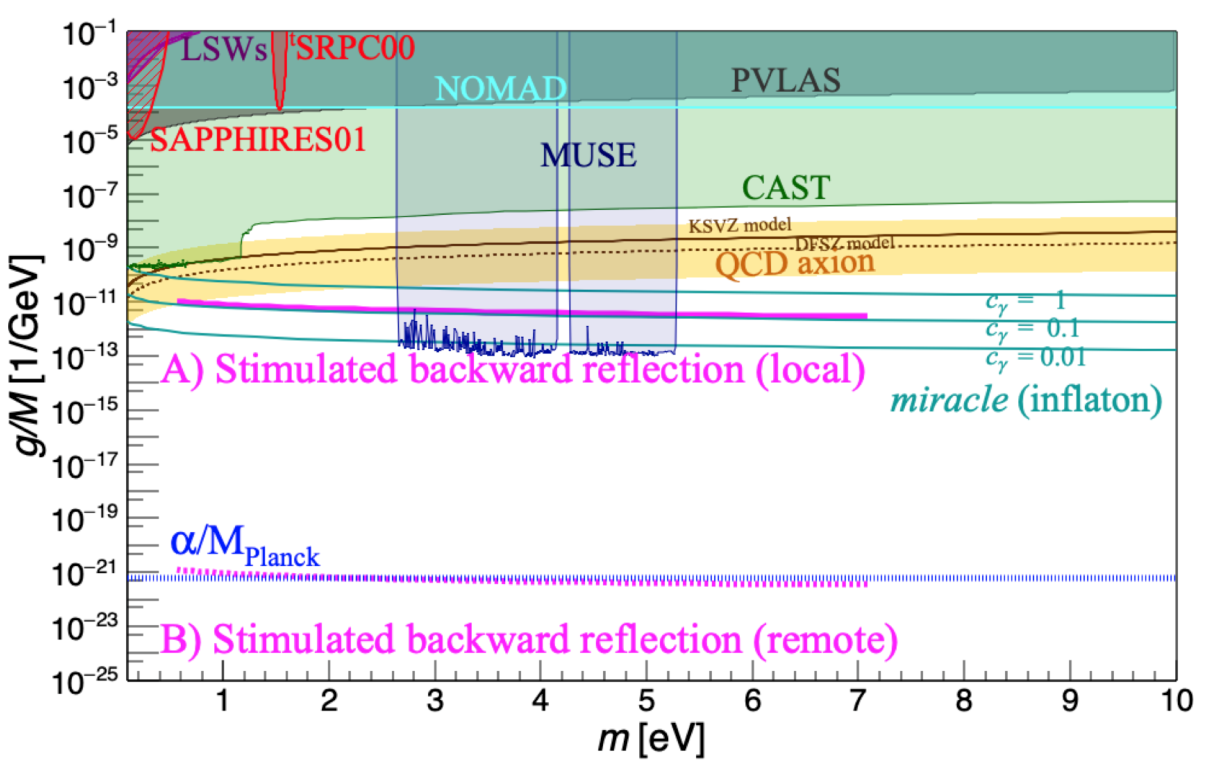}
\caption{
Sensitivity projections based on backward reflection from stimulated axion
decay by surveying with a spherical inducing laser field.
The details of symbols and used parameters can be found in the main text.
}
\label{Fig2}
\end{center}
\end{figure}
 
The magenta solid and dashed horizontal curves in Fig.\ref{Fig2} show 
expected reachable coupling strengths
obtained from Eq.(\ref{eq_goM}) for A) and B), respectively,
as a function of axion mass $m=2\om_i$ covered by the laser spectrum range.
For simplicity we commonly assume the inducing laser pulse energy of 1~J
and the divergence angle at the incident points, $\theta_0 = 1$~mrad.
The range of photon wavelengths is $\lambda_i=0.35-4.4 \quad\mu$m.
The searching depths of $L_0=0 \sim L_1=10^3$~m and
$L_0=0 \sim L_1=F(R_E)-F(0)= 2.3 \times 10^9$~m are assumed
for A) and B), respectively.
The upper solid brown line and yellow band are the predictions from the benchmark QCD axion model:
the KSVZ model~\cite{KSVZ} with the model parameters $E/N = 0$ and 
with $0.07 < \left|E/N - 1.95\right| < 7$, respectively, 
while the bottom dashed brown line is the prediction from the DFSZ model~\cite{DFSZ} with $E/N = 8/3$.
The cyan lines are the predictions from the ALP {\it miracle} model relevant to inflation\cite{MIRACLE}
with the model parameters $c_{\gamma}=1, 0.1, 0.01$.
The gray area shows the excluded region by the vacuum magnetic birefringence experiment (PVLAS~\cite{PVLAS}).
The purple areas are excluded regions by the Light-Shining-through-a-Wall (LSW) experiments
(ALPS~\cite{ALPS} and OSQAR~\cite{OSQAR}).
The light-cyan horizontal solid line indicates the upper limit from the
search for eV (pseudo)scalar penetrating particles in the SPS neutrino beam (NOMAD)~\cite{NOMAD}.
The blue areas are exclusion regions from the optical MUSE-faint survey~\cite{MUSE}.
The green area indicates the excluded region by the helioscope experiment CAST \cite{CAST}.
The red shaded area shows the excluded range based on a stimulated resonant photon collider, 
SRPC, in quasi-parallel collision geometry (SAPPHIRES01)~\cite{SAPPHIRES01}.
The red filled area indicates the excluded range by a pilot search 
with the fixed angle three-beam SRPC(\tSRPC00)~\cite{3beam01}.

\section{Conclusion}
In conclusion,
we have formulated the stimulated decay process of a pseudoscalar particle 
into two photons 
in the background of an inducing coherent electromagnetic field and applied the formula to 
obtain the stimulated backward reflection yield based on the spherical wave propagation of 
the inducing laser field.
For the laboratory based experiment targeting locally moving cold dark matter,
the sensitivity can reach $f_a \sim 10^7$~GeV, for instance, 
based on the concrete relation $g/M=\alpha/(2\pi f_a)C_{\gamma}=1.2 \times 10^{-11}$~GeV${}^{-1}$
with $C_{\gamma}=0.1$ in the model~\cite{Badziak:2023fsc}.
We also have evaluated the yield 
if we apply this method to the dark matter components in the S1 stream from Cygnus.
We considered the case where a satellite trailing the Earth's orbit constantly faces
the S1 stream so that the position of the dark matter hair-root
can be fixed with respect to the detector in the satellite. 
The 4th power dependence on the search depth of the signal yield 
can significantly improve the sensitivity to the weak coupling domain. 
This evaluation indeed indicates very high sensitivity reaching the gravitational
coupling strength $\alpha/M_{Planck} = 6 \times 10^{-22}$~GeV${}^{-1}$.

We have focused on the eV mass range as the concrete application of the proposed method
in this paper, 
however, this method is in principle extendable to any frequency ranges of inducing beams. 
Given intense coherent photon sources spanning many orders of magnitude in frequency, 
laboratory searches on the ground are quite feasible provided a vacuum space of km size 
is available.
In practice, ground based experiments have advantage in capability of supplying high power
to coherent electromagnetic sources.
This method can also provide a measure for transverse velocity distributions of 
local dark matter based on divergence angles of reflected signal photons. 
Once the mass is identified through signs of dark matter observation on the ground, 
the search for hair-roots in space using an inducing field with a specific wavelength
consistent with the mass would become a realistic option.

\section*{Acknowledgments}
The author thanks Tomohiro Inagaki, Taiyo Nakamura and Tsunefumi Mizuno 
for useful discussions on stimulated yield for cold dark matter, 
Kazunori Kohori on the concentration factor around the surface of Earth, 
and Kazuhisa Mitsuda on possible satellite orbits in space.
The author also acknowledges the support of the Collaborative Research
Program of the Institute for Chemical Research at Kyoto University
(Grant Nos.\ 2023--101 and 2024--95)
and Grants-in-Aid for Scientific Research Nos.\ 21H04474 from the Ministry of Education,
Culture, Sports, Science and Technology (MEXT) of Japan.

\end{document}